\documentstyle[]{mn}
\newif\ifAMStwofonts

\def\etal{{\rm et al.}}

\def\simgt{\mathrel{\spose{\lower 3pt\hbox{$\sim$}}
        \raise 2.0pt\hbox{$>$}}}
\def\simlt{\mathrel{\spose{\lower 3pt\hbox{$\sim$}}\raise 2.0pt\hbox{$<$}}}

\ifoldfss
  \ifCUPmtlplainloaded \else
    \NewTextAlphabet{textbfit} {cmbxti10} {}
    \NewTextAlphabet{textbfss} {cmssbx10} {}
    \NewMathAlphabet{mathbfit} {cmbxti10} {} 
    \NewMathAlphabet{mathbfss} {cmssbx10} {} 
  \fi
  \ifAMStwofonts
    \ifCUPmtlplainloaded \else
      \NewSymbolFont{upmath} {eurm10}
      \NewSymbolFont{AMSa} {msam10}
      \NewMathSymbol{\upi}     {0}{upmath}{19}
      \NewMathSymbol{\umu}     {0}{upmath}{16}
      \NewMathSymbol{\upartial}{0}{upmath}{40}
      \NewMathSymbol{\leqslant}{3}{AMSa}{36}
      \NewMathSymbol{\geqslant}{3}{AMSa}{3E}

    \fi
  \fi
\fi 

\ifnfssone
  \newmathalphabet{\mathit}
  \addtoversion{normal}{\mathit}{cmr}{m}{it}
  \addtoversion{bold}{\mathit}{cmr}{bx}{it}
  \newmathalphabet{\mathbfit} 
  \addtoversion{normal}{\mathbfit}{cmr}{bx}{it}
  \addtoversion{bold}{\mathbfit}{cmr}{bx}{it}
  \newmathalphabet{\mathbfss} 
  \addtoversion{normal}{\mathbfss}{cmss}{bx}{n}
  \addtoversion{bold}{\mathbfss}{cmss}{bx}{n}
  \ifAMStwofonts
    \ifCUPmtlplainloaded \else
      \UseAMStwoboldmath
      \makeatletter
      \new@mathgroup\upmath@group
      \define@mathgroup\mv@normal\upmath@group{eur}{m}{n}
      \define@mathgroup\mv@bold\upmath@group{eur}{b}{n}
      \edef\UPM{\hexnumber\upmath@group}
      \new@mathgroup\amsa@group
      \define@mathgroup\mv@normal\amsa@group{msa}{m}{n}
      \define@mathgroup\mv@bold\amsa@group{msa}{m}{n}
      \edef\AMSa{\hexnumber\amsa@group}
      \makeatother
      \mathchardef\upi="0\UPM19
      \mathchardef\umu="0\UPM16
      \mathchardef\upartial="0\UPM40
      \mathchardef\leqslant="3\AMSa36
      \mathchardef\geqslant="3\AMSa3E
    \fi
  \fi
\fi 

\ifnfsstwo
  \DeclareMathAlphabet{\mathbfit}{OT1}{cmr}{bx}{it}
  \SetMathAlphabet\mathbfit{bold}{OT1}{cmr}{bx}{it}
  \DeclareMathAlphabet{\mathbfss}{OT1}{cmss}{bx}{n}
  \SetMathAlphabet\mathbfss{bold}{OT1}{cmss}{bx}{n}
  \ifAMStwofonts
    \ifCUPmtlplainloaded \else
      \DeclareSymbolFont{UPM}{U}{eur}{m}{n}
      \SetSymbolFont{UPM}{bold}{U}{eur}{b}{n}
      \DeclareSymbolFont{AMSa}{U}{msa}{m}{n}
      \DeclareMathSymbol{\upi}{0}{UPM}{"19}
      \DeclareMathSymbol{\umu}{0}{UPM}{"16}
      \DeclareMathSymbol{\upartial}{0}{UPM}{"40}
      \DeclareMathSymbol{\leqslant}{3}{AMSa}{"36}
      \DeclareMathSymbol{\geqslant}{3}{AMSa}{"3E}
    \fi
  \fi
\fi 

\ifCUPmtlplainloaded \else
  \ifAMStwofonts \else 
    \def\upi{\pi}
    \def\umu{\mu}
    \def\upartial{\partial}
  \fi
\fi

\title[The microlens mass function of Q2237+0305]
  {Limits on the microlens mass function of Q2237+0305}
\author[J. S. B. Wyithe et al.]
  {J.~S.~B.~Wyithe$^1$, 
  R.~L.~Webster$^1$,
  E. L.~Turner$^2$\\
  $^1$ School of Physics, University of Melbourne, Parkville, Vic, 3052, 
Australia\\
  $^2$ Princeton University Observatory, Peyton Hall, Princeton, NJ 08544, USA\\ 
 Email: swyithe@physics.unimelb.edu.au, rwebster@physics.unimelb.edu.au, elt@astro.princeton.edu }
\date{Accepted Received}
\pagerange{\pageref{firstpage}--\pageref{lastpage}}
\pubyear{1999}

\def\LaTeX{L\kern-.36em\raise.3ex\hbox{a}\kern-.15em
    T\kern-.1667em\lower.7ex\hbox{E}\kern-.125emX}

\begin{document}

\label{firstpage}

\maketitle

\begin{abstract}

Gravitational microlensing at cosmological distances is potentially a powerful tool for probing the mass functions of stars and compact objects in other galaxies. In the case of multiply-imaged quasars, microlensing data has been used to determine the average microlens mass. However the measurements have relied on an assumed transverse velocity for the lensing galaxy.
Since the measured mass scales with the square of the transverse velocity, published mass limits are quite uncertain. In the case of Q2237+0305 we have properly constrained this uncertainty. The distribution of light curve derivatives allows quantitative treatment of the relative rates of microlensing due to proper motions of microlenses, the orbital stream motion of microlenses and the bulk galactic transverse velocity. By demanding that the microlensing rate due to the motions of microlenses is the minimum that should be observed we determine lower limits for the average mass of stars and compact objects in the bulge of Q2237+0305. If microlenses are assumed to move in an orbital stream the lower limit ranges between 0.005 and 0.023$M_{\odot}$ where the the systematic dependence is due to the fraction of smooth matter and the size of photometric error assumed for published monitoring data. However, if the microlenses are assumed to move according to an isotropic velocity dispersion then a larger lower limit of 0.019-0.11$M_{\odot}$ is obtained. A significant contribution of Jupiter mass compact objects to the mass distribution of the galactic bulge of Q2237+0305 is therefore unambiguously ruled out.

\end{abstract}


\section{Introduction}

Q2237+0305 comprises a source quasar with a redshift of $z=1.695$ that is gravitationally lensed by a foreground galaxy at $z=0.0394$ producing 4 images with separations of $\sim 1''$.  Each of the 4 images are observed through the bulge of a galaxy which has an optical depth in stars that is of order unity (eg. Kent \& Falco 1988; Schneider et al. 1988; Schmidt, Webster \& Lewis 1998). In addition, the proximity of the lensing galaxy means that the effective transverse velocity may be high. The combination of these facts make Q2237+0305 the ideal object from which to study microlensing. Indeed, Q2237+0305 is the only object in which cosmological microlensing has been confirmed (Irwin et al. 1989; Corrigan et al. 1991).

Numerical microlensing simulations (eg. Wambsganss, Paczynski \& Schneider 1990; Witt, Kayser \& Refsdal 1993) have shown that the statistics of microlensed high magnification events (HMEs) obtained from long term monitoring may provide information on properties of the lens such as the stellar mass function and the percentage of mass in stars for the bulge. However, the monitoring period required is greater than 100 years (Wambsganss, Paczynski \& Schneider 1990). There are several other unknown quantities in the problem. These include the magnitude and direction of any transverse motion, as well as the source size. In particular, the magnitude and direction of the transverse motion are degenerate with the density of caustics (a function of the mean compact object mass) for a given set of microlensing statistics.

 Microlensed fluctuation in the quasars continuum results from motion due to both a galactic transverse velocity and to the random proper motion or stream motion of stars and compact objects. Wyithe, Webster \& Turner (1999a) (hereafter  WWTa) define the equivalent transverse velocity as the transverse velocity in a model containing stars with static positions, that produces a rate of microlensing most closely resembling that of a model containing stellar proper motions. Foltz et al. (1992) measured the central velocity dispersion of Q2237+0305 to be $\sim$ 215$km\,sec^{-1}$, and theoretical models (Schmidt, Webster \& Lewis 1998) predict a value of $\sim$ 165$km\,sec^{-1}$. If the dispersion is isotropic then (as discussed in Wyithe, Webster \& turner (1999b) (hereafter  WWTb)) the equivalent transverse velocity calculated from the overall microlensing rate (all 4 images) of Q2237+0305 is larger than this value. The microlensing rate that results from the line-of-sight velocity dispersion (of an isotropic distribution) is therefore comparable to that of the likely transverse velocity. This means that the often made assumption that random proper motions provide a negligible contribution to microlensing in Q2237+0305 is incorrect.

While the microlensing rate resulting from random stellar motions is dependent on the mean microlens mass, the equivalent transverse velocity (WWTa; WWTb) that describes the microlensing rate is not. We use this fact to break the degeneracy between microlensing rate and mean microlens mass, and hence obtain useful limits on the mass function along the line-of-sight through the bulge of the lensing galaxy.

This paper is presented in 5 parts. Sec.~\ref{mass} contains a general discussion of microlensing and the mass function. Sec.~\ref{models} discusses the numerical methods used to model microlensing in Q2237+0305 and Sec.~\ref{mass_limits} describes a method to place limits on the mass function through consideration of microlensing due to both a transverse velocity and a stellar velocity dispersion.

\section{Microlensing and the stellar mass function}
\label{mass}

\begin{table*}
\begin{center}
\caption{\label{ml}A collection of results from microlensing data for the average mass of compact objects in galactic halos and/or bulges of the Milky Way and Quasar lensing galaxies.}
\begin{tabular}{|c|c|c|}
\hline

Milky Way halo & Results for mean masses of dark halo objects toward LMC.  &   $\langle m\rangle=0.13^{+.08}_{-.05}M_{\odot}$\\
(Alcock et al. 1997a) & The ranges include statistical ($1\sigma$) as well as systematic &\hspace{7mm} -$0.55^{+.38}_{-.21}M_{\odot}$   \\
                            & uncertainties due to choice of halo model. & \vspace{3mm}\\

Milky Way bulge & Microlensing event time-scales are consistent with a mean & $0.1M_{\odot}\la\langle m\rangle\la1.0M_{\odot} $ \\
(Alcock et al. 1997b) & mass of compact objects and stars in the Milky Way bulge.\vspace{3mm}\\

0957+561 halo & The limit has 99\% and 95\% significance levels for assumed & \\
(Schmidt \& Wambsganss 1998) &   source sizes of $4\times10^{14}cm$ and $4\times10^{15}cm$, and scales as & $\langle m\rangle \ga 0.001\left(\frac{v_{t}}{600}\right)^{2}$ \\
	      & the square of the assumed transverse velocity $v_{t}$. & \vspace{3mm}\\
		
2237+0305 bulge and halo & The microlensing rate of current light curves is consistent & \\
 (Lewis \& Irwin 1996) & with a mean mass of stars and compact objects. The value & $0.1M_{\odot}\la\langle m\rangle\left(\frac{v_{t}}{600}\right)^{2} \la10M_{\odot} $\\
                       & scales as the square of the assumed transverse velocity $v_{t}$. &  \vspace{3mm}\\

2237+0305 bulge and halo & The proper motions of microlenses provide a minimum micro- &  $\langle m\rangle = 0.010^{+?}_{-.005}M_{\odot}$ \\
This paper.		& lensing rate. The limits given for the mean mass have a 99\% & \hspace{7mm}$-0.29^{+?}_{-.18}M_{\odot}$ \\
                        & significance level, and depend on the bulge dynamics assumed.   & \\\hline

\end{tabular}
\end{center}
\end{table*}

Stellar mass functions have traditionally been measured through the combination of an observed luminosity function and an empirical mass-luminosity relationship. However, because of the difficulties inherent in the observations of faint stars, these determinations become uncertain as the hydrogen burning limit ($\sim 0.08M_{\odot}$) is approached and crossed. In contrast to this approach, microlensing uses the mass of a lens to magnify the observed flux of a background source through gravitational deflection of the light bundle. The statistics obtained are therefore free of any bias introduced by the hydrogen burning limit, and so microlensing is a powerful tool for determining the contribution to the mass function of low mass stars and dark compact objects. 

Gravitational microlensing is observed in two very different regimes. Paczynski (1986) suggested surveying millions of stars in the Magellanic Clouds for microlensing induced flux variation as a way of detecting solar mass compact objects in the halo of our own galaxy. Several groups have undertaken such searches (eg. Alcock et al. 1997a; Renault et al. 1998). In a similar vein, Paczynski (1991) and Griest et al. (1991) suggested microlensing experiments towards the Galactic bulge as a method to probe the masses of stars along the line-of-sight in the Galactic disc. Microlensing searches towards the Galactic bulge have since found a higher rate of microlensing events than are found towards the LMC (eg. Alcock et al. 1997b; Udalski et al. 1994). In a very different microlensing regime, the observed flux of a back-ground quasar can be altered by gravitational lensing by stars or halo objects in a foreground galaxy. Q2237+0305 is an example of a quasar that lies very close to a galactic line-of-sight. In such cases multiple imaging occurs, allowing microlensed variation to be easily separated from intrinsic fluctuations which are observed in all images.

 The analyses of microlensing in the Galactic and cosmological regimes are very different. While there are only a few sources (quasar images) in the cosmological case rather than the millions available in Galactic microlensing searches, the optical depth (or equivalently the probability of lensing) is $\sim 10^{6}$ greater. The transverse velocity is a critical parameter for the determination of a mass or masses responsible for a microlensing event, regardless of the microlensing scenario. Unfortunately it is unknown in both cases. In Galactic microlensing calculations the velocities are inferred from an assumed distribution. However in the cosmological case the transverse velocities of the lensing objects are not independent. Rather, they are equal except for the contribution of the individual stellar proper motions. Moreover, cosmological microlensing results from the gravitational contribution of a large ensemble of many hundreds of masses rather than on a single microlens. Microlensed, multiply-imaged quasars are an excellent probe of the mass function of compact objects along the quasar image line-of-sight because they interact with a large number of microlenses.

Several authors have placed limits on the masses of microlenses responsible for cosmological microlensing. Schmidt \& Wambsganss (1998) use the lack of observed variation in Q0957+561 to place a lower limit on the mass of microlenses in the halo of the lensing galaxy. They rule out a mean mass of halo objects $\langle m\rangle\ll 10^{-2}M_{\odot}$. However their determination is dependent on the source size, and the fraction of smooth matter employed in their calculations. A transverse velocity of $v_{t}=600\,km\,sec\,^{-1}$ is assumed. The uncertainty in this assumed value is the most serious problem for this kind of analysis because the values of mass obtained are $\propto v_{t}^{2}$. Lewis \& Irwin (1996) compared the monitoring data of Q2237+0305 with simulations using a structure function to analyse variability. They too assume a transverse velocity of $v_{t}=600\,km\,sec\,^{-1}$ and conclude that the mean mass of objects in Q2237+0305 is $0.1M_{\odot}\la\langle m\rangle \la 10M_{\odot}$. This result is also proportional to the square of the unknown transverse velocity. 

While they do not measure precisely the same quantity, the above results are supportive of those of the MACHO experiment (eg Alcock et al. 1997a,b). From microlensing observations towards the LMC by the Galactic halo, Alcock et al. (1997a) conclude that the average MACHO masses are $0.08-0.93M_{\odot}$. The quoted range includes both statistical uncertainties and systematic effects introduced by the assumption of halo model. In addition, they find evidence for an excess of events over those predicted from stellar lensing alone. Alcock et al. (1997b) find that the range of time-scales towards the Galactic bulge are consistent with microlens masses of $0.1M_{\odot}\la\langle m\rangle \la 1.0M_{\odot}$. Tab.~\ref{ml} summarises the results described.

\section{The microlensing models}
\label{models}
\subsection{Microlensing models for Q2237+0305}

\begin{table}
\begin{center}
\caption{\label{params}Values of the total optical depth and the magnitude of the shear at the position of each of the 4 images of Q2237+0305. The quoted values are those of Schmidt, Webster \& Lewis (1998). $\kappa_{*}$ and $\kappa_{c}$ are the optical depths in stars and in smoothly distributed matter respectively.}
\begin{tabular}{|c|c|c|}
\hline
Image & $\kappa=\kappa_{*}+\kappa_{c}$   & $|\gamma|$  \\ \hline
  A   & 0.36                             &  0.40        \\
  B   & 0.36                             &  0.40        \\
  C   & 0.69                             &  0.71        \\
  D   & 0.59                             &  0.61        \\ \hline
\end{tabular}
\end{center}
\end{table}

 Throughout the paper, standard notation for gravitational lensing is used. The Einstein radius of a 1$M_{\odot} $ star in the source plane is denoted by $\eta_{0} $. The normalised shear is denoted by $\gamma$, and the convergence or optical depth by $\kappa$. The model for gravitational microlensing consists of a very large sheet of point masses that simulates the section of galaxy along the image line-of-sight, together with a shear term that includes the perturbing effect of the mass distribution of the lensing galaxy as a whole. The normalised lens equation for a field of point masses with an applied shear in terms of these quantities is
\begin{equation}
\vec{y}= \left( \begin{array}{cc}
	 1-\kappa_{c}-\gamma & 0 \\
	0 & 1-\kappa_{c}+\gamma 
	    \end{array} \right)\vec{x} + \sum_{j=0}^{N_{*}}m^{j}\frac{(\vec{x}^{j}-\vec{x})}{|\vec{x}^{j}-\vec{x}|^{2}}
\label{lens_map} 
\end{equation}
Here $\vec{x}$ and $\vec{y}$ are the normalised image and source positions respectively, and the $\vec{x}_{i}^{j}$ and $m^{j}$ are the normalised positions and masses of the microlenses. $\kappa_{c}$ is the optical depth in smoothly distributed matter. Eqn. \ref{lens_map} is solved for the macroimage magnification at many points along a predefined source trajectory through the inversion technique of Lewis et al. (1993) and Witt (1993). The region of the lens plane in which image solutions need to be found to ensure that $99\%$ of the total macro-image flux is recovered from a source point was described by Katz, Balbus \& Paczynski (1986). The union of areas of in the lens plane corresponding to the flux collection area of each point on the source line is known as the shooting region, the method for determining the dimensions of which is described in Lewis \& Irwin (1995), and Wyithe \& Webster (1999). The radius of the disc of point masses is chosen to be 1.2 times that required to cover this shooting region.

We assume the macro-parameters for Q2237+0305 calculated by Schmidt, Webster \& Lewis (1998) (the values are shown in Tab.~\ref{params}). Two models are considered for the mass distribution, one with no continuously distributed matter and one where smooth matter contributes 50\% of the surface mass density. Both the microlensing rate due to a transverse velocity (Witt, Kayser \& Refsdal 1993; Lewis \& Irwin 1996; WWTb), as well as the corresponding rate due to proper motions (WWTa) are not functions of the detail of the microlens mass distribution, but depend only on the mean microlens mass.
The mean of the microlens mass function can therefore be determined independently from its form. On the other hand, information on the form of the mass function is unobtainable through consideration of the microlensing rate. We limit our attention to models in which all the point masses have identical mass.

\begin{figure}
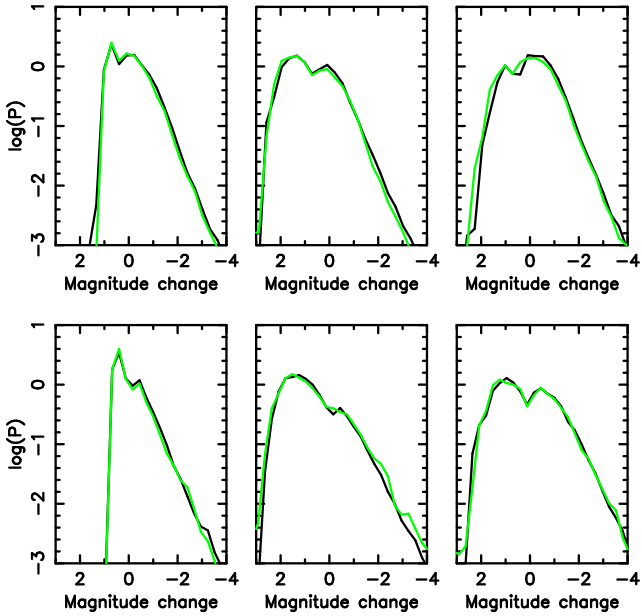

\vspace*{85mm}
\includegraphics{fig1a.epsi}
\includegraphics{fig1b.epsi}
\caption{\label{amp_dist} The magnification distributions for images A/B (left), C (centre) and D (right). The top and bottom rows show the distributions for models that have no smooth matter component, and a 50\% smooth matter component respectively. The light and dark lines represent distributions that are sampled parallel to and at right angles to the shear.}
\end{figure}

 Each of our models was computed for a source track of length 10$\eta_{o}$ (corresponding to $\sim70$ years for an effective transverse velocity of $600\,km\,sec^{-1}$). We set 500 to be the minimum number of stars in a model. Two orientations were chosen for the transverse velocity with respect to the galaxy, with the source trajectory being parallel to the A$-$B or C$-$D axes. At each orientation, 100 simulations were made for each of the 4 images of Q2237+0305 in combination with each of the model mass distributions. The two orientations bracket the range of possibilities, and because the images are positioned approximately orthogonally with respect to the galactic centre correspond to shear values of $\gamma_{A},\gamma_{B}<0,\gamma_{C},\gamma_{D}>0$ and $\gamma_{A},\gamma_{B}<0,\gamma_{C},\gamma_{D}>0$ respectively. Tab.~\ref{param} shows the number of stars used for each of these models along with the mean magnification $\langle \mu \rangle$, and the theoretical magnification $\langle \mu_{th}\rangle$ for comparison. The average magnification was found from the combination of the two sets of models, and the error calculated from the standard deviation of a subset of 6 simulations (3 per model orientation). Fig.~\ref{amp_dist} shows the magnification distributions for each image in each model. Simulations having $\gamma<0$ and $\gamma>0$ are represented by light and dark lines respectively. The distributions are quantitatively comparable to those presented for similar values of $\kappa$ and $\gamma$ in Lewis \& Irwin (1995), and demonstrate that the magnification distribution is independent of the source direction as required.

\begin{table}
\begin{center}
\caption{\label{param}Values for the number of stars in the microlensing models (left column $\gamma>0$; right column $\gamma<0$), the average model magnifications and the theoretical values for comparison.}
0\% smooth matter
\begin{tabular}{|c|c|c|c|c|}
\hline
Image & \multicolumn{2}{c}{No.\ Stars}  &  $\langle \mu \rangle$  &  $\langle \mu_{th} \rangle$ \\ \hline\hline
  A/B & 763 &  500     &  3.97$\pm$.28           &        4.01                 \\    
  C   & 1022 &  500    &  2.39$\pm$.11           &        2.45                 \\ 
  D   & 2887 &  1301   &  4.79$\pm$.14           &        4.90                 \\ \hline \\\\
\end{tabular}

50\% smooth matter
\begin{tabular}{|c|c|c|c|c|}
 \hline
Image &  \multicolumn{2}{c}{No.\ Stars}  &  $\langle \mu \rangle$  &  $\langle \mu_{th} \rangle$ \\ \hline\hline
  A/B   & 500 & 500   &  4.01$\pm$.19           &        4.01                 \\
  C   & 500 & 500     &  2.44$\pm$.09           &        2.45                 \\
  D   & 1080 & 500    &  4.84$\pm$.20           &        4.90                 \\ \hline

\end{tabular}
\end{center}
\end{table}

We assume that microlensing is produced through the combination of a galactic transverse velocity with each of two classes of proper motion for individual stars with respect to the galaxy: an isotropic velocity dispersion and a circular stream motion. Moreover, we assume that the magnitude of the dispersion or stream motions are the same for each of the four images. We take the theoretical value of $\sigma_{*}\sim 165\,km\,sec^{-1}$ for the line-of-sight velocity dispersion of the stars in the galactic bulge. This is lower than the value observed by Foltz et al. (1992), so lower limits placed on the mass are more conservative.

\subsection{The effective transverse velocity}

\begin{figure}
\vspace*{80mm}
\includegraphics{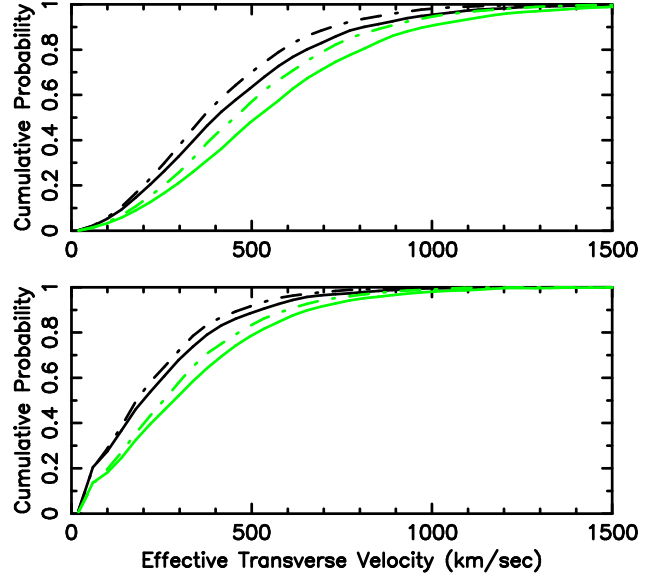}
\caption{\label{vel_lim} Plots of the cumulative probability ($P_v$) for the effective transverse velocity. The solid and dot-dashed lines correspond to the models with trajectory directions that are aligned with the C-D and A-B image axes respectively. The dark and light lines represent results from models containing 0\% and 50\% continuously distributed matter. The upper and lower plots correspond to the cases where the photometric error was assumed to be $\sigma_{SE}$=0.01 mag in images A/B and $\sigma_{SE}$=0.02 in images C/D, and $\sigma_{LE}$=0.02 mag in images A/B and $\sigma_{LE}$=0.04 in images C/D. The model assumed $1M_{\odot}$ microlenses.}
\end{figure}

We define the effective transverse velocity as the transverse velocity that produces a microlensing rate from a static field model equal to that of the observed light-curve. The effective transverse velocity therefore describes the microlensing rate due to the combination of the effects of a galactic transverse velocity and random microlens proper motion.

Our calculation of effective transverse velocity utilises the cumulative histogram of derivatives calculated from the 6 difference light-curves (A-B, A-C, A-D, B-C, B-D, C-D). Through computation of the difference light-curves, intrinsic flux variation, as well as the systematic component of the observational uncertainty are removed from the data. WWTb describes a procedure for determining the probability that the effective transverse velocity is less than an assumed value. That paper provides correct upper limits, but does not obtain the cumulative probability function. A modified procedure to determine the cumulative probability for effective transverse velocity from ensembles of mock data-sets is described below. 

Many mock observations of the 6 difference light-curves were produced from model light-curves with a sampling rate and period that are identical to the published monitoring data (Irwin et al. 1989, Corrigan et al. 1991, $\O$stensen et al. 1996). Observational errors were also simulated by applying a random fluctuation to each point on the light curve, distributed according to a Gaussian with a half-width $\sigma$ describing the published photometric error.
 The simulations used two different estimates for the uncertainty in the photometric magnitudes. In the first case a small error was assumed (SE). For images A and B, $\sigma_{SE}$=0.01 mag, and for images C and D $\sigma_{SE}$=0.02 mag. In the second case, a larger error was assumed (LE). For images A and B, $\sigma_{LE}$=0.02 mag and for images C and D $\sigma_{LE}$=0.04 mag. The observational error in Irwin et al. (1989) was 0.02 mag. We note that the adoption of a larger error leads to a smaller bound on the measured upper limit of effective transverse velocity, and subsequently a larger lower bound on the mean mass.

The simulations assume a point source since the value of effective transverse velocity measured with a model is not a sensitive function of the source size (in the range of interest) assumed (WWTb). The insensitivity arises from the fact that the measurement is derived predominantly from points that are not part of a HME.

We define a set of effective transverse velocities ($V_{eff}$) and determine the probability, based on light-curve derivatives that each of these correctly describes the true value for Q2237+0305 ($V_{gal}$). For each model, 5000 mock observations were produced at each effective transverse velocity from the four sets (one per image) of 100 10$\eta_{o}$ simulated light-curves. Due to the finite sampling period, histograms of microlensed light-curve derivatives produced from the individual mock observations describe typical but not average behaviour. Therefore an average histogram was also computed from each set of mock observations. For each mock observation at each pre-defined effective transverse velocity ($V_{eff}$), a mock measurement of effective transverse velocity $v_{eff}$ was made by minimising the KS difference between the average histogram for $v_{eff}$ and the histogram of the mock observation. Thus we calculate the function representing the likelihood $p_{lhood}(v_{eff}|V_{eff})$ for observing $v_{eff}$ given an assumption for the true value ($V_{eff}$).

Similarly, from the observed histogram of microlensed light-curve derivatives we find the effective transverse velocity ($v_{obs}$) that best describes the observed microlensing rate. Using Bayes' theorem we calculate the posterior probability that the effective galactic transverse velocity $V_{gal}$ is less than an assumed value $V_{eff}$: 
\begin{eqnarray}
\nonumber P_{v}(V_{gal}<V_{eff}|v_{obs})&=&\\
&&\hspace{-25mm}\int_{0}^{V_{eff}}p_{lhood}(v_{obs}|V_{eff}')\,p_{prior}(V_{eff}')\,dV_{eff}',
\end{eqnarray} 
where $p_{prior}(V_{eff})$ is the prior probability for $V_{eff}$.
We have assumed two different and physically un-motivated priors. These are respectively flat:
\begin{eqnarray}
\nonumber p_{prior}(V_{eff})&\propto& dV_{eff}\hspace{10mm} 0<V_{eff}<2000\sqrt{\langle m\rangle}\\
p_{prior}(V_{eff})&\equiv&0\hspace{16mm} otherwise, 
\end{eqnarray}
and logarithmic: 
\begin{eqnarray}
\nonumber p_{prior}(V_{eff})&\propto&\frac{dV_{eff}}{V_{eff}}\hspace{10mm} 0<V_{eff}<2000\sqrt{\langle m\rangle}\\
p_{prior}(V_{eff})&\equiv&0\hspace{16mm} otherwise. 
\end{eqnarray}
These priors bracket the range and spread of physically plausible priors such as a Gaussian based on observed peculiar velocities (eg Mould et al. 1993) or those computed from numerical studies. Our results are insensitive to the choice of these priors, and we conclude that our estimate of probability is dominated by the microlensing observations rather than our assumed prior for $V_{eff}$.

$P_v(v_{gal}<V_{eff}|v_{obs})$ is plotted in Fig.~\ref{vel_lim} for the models discussed in this paper ($\langle m\rangle = 1M_{\odot}$). In these plots the solid and dot-dashed lines correspond to source trajectories along the C-D image axis and A-B image axis, and the dark and light lines correspond to models with 0\% and 50\% smoothly distributed matter. The upper and lower plots correspond to the cases where the photometric error was assumed to be $\sigma_{SE}$=0.01 mag in images A/B and $\sigma_{SE}$=0.02 in images C/D, and $\sigma_{LE}$=0.02 mag in images A/B and $\sigma_{LE}$=0.04 in images C/D.

\subsection{The contribution to microlensing of stellar proper motions}

The average histogram of light-curve derivatives can be computed for the case where microlensing results from a random velocity dispersion rather than a transverse velocity. We define the equivalent transverse velocity as the transverse velocity that in combination with a static microlensing model produces a microlensing rate closest to the rate produced by a random velocity dispersion of the point masses. The microlensing rates are considered equivalent at the transverse velocity that produces a cumulative histogram of light-curve derivatives closest (has the minimum possible KS difference) to the corresponding proper motion histogram.

 The formalism required for the computation of the cumulative distribution of derivatives resulting from proper motion of stars was developed in WWTa. The upper panel of Tab.~\ref{stream_tab} shows the values of equivalent transverse velocity of a Gaussian 1-d velocity dispersion with a half-width of $\sigma_{*}=165\,km\,sec^{-1}$. These values were computed from the difference light-curves for models of Q2237+0305 (the quoted error describes the range of values obtained over three separate sets of simulations). The two cases shown are for a trajectory that is parallel to the shear in images A and B ($\gamma_{A},\gamma_{B}>0$), and for one that is parallel to the shear in images C and D ($\gamma_{C},\gamma_{D}>0$). In all cases the effective transverse velocity is larger than the 1-d velocity dispersion.

\section{The mass of compact objects in Q2237+0305}
\label{mass_limits}

\subsection{Placing limits on the mass using effective transverse velocity}

\begin{table*}
\begin{center}
\caption{\label{stream_tab}Table showing the equivalent transverse velocities and the 99\%, 95\% and  90\% lower limits for the mean microlens mass as well as the most likely value. Values for the mass limits are shown corresponding to the cases where the error was $\sigma_{SE}=0.01$ mags in images A/B, $\sigma_{SE}=0.02$ mags in images C/D, and where the error was  $\sigma_{LE}=0.02$ mags in images A/B, $\sigma_{LE}=0.04$ mags in images C/D (in parentheses). Values are shown that correspond to an isotropic velocity dispersion of $\sigma{*}=165\,km\,sec^{-1}$ (top table), a circular stream velocity of $v_{stream}=233\,km\,sec^{-1}$ (bottom table).}

\begin{center}

\vspace{5mm}

 $\sigma{*}=165\,km\,sec^{-1}$
\begin{tabular}{|c|c|c|c|c|c|c|c|}

   \hline
 Trajectory                                                  & Smooth  & Equiv.           & Bayesian      & $m_{low}(99\%)$  & $m_{low}(95\%)$   & $m_{low}(90\%)$  &  mode          \\
Orientation                                                  & Matter  & Vel. ($km\,sec^{-1}$)        &  prior       &   ($M_{\odot}$)  &  ($M_{\odot}$)    &  ($M_{\odot}$)    &  ($M_{\odot}$)     \\ \hline\hline 
Shear:                                                       &  0\%    & 280 $\pm$15      &   log        &   0.050 (0.111)    &  0.111 (0.277)      &  0.176 (0.484)    &  0.139 (0.288)     \\
 $\gamma_{A},\gamma_{B}>0,\:\:\gamma_{C},\gamma_{D}<0  $     &  50\%   & 199 $\pm$12      &   log        &   0.019 (0.039)    &  0.042 (0.095)      &  0.066 (0.161)    & 0.054  (0.087)    \vspace{3mm}
 \\

Shear:                                                       &  0\%    & 270 $\pm$10      &   log        &    0.053 (0.111)   & 0.111 (0.270)       & 0.173 (0.463)     & 0.139 (0.281)      \\
$\gamma_{A},\gamma_{B}<0,\:\:\gamma_{C},\gamma_{D}>0  $      &  50\%   & 181 $\pm$4       &   log        &    0.018 (0.036)   & 0.039 (0.084)        & 0.061 (0.143)     & 0.042 (0.087)     \\ \hline

Shear:                                                       &  0\%    & 280 $\pm$15      &   flat       &    0.098 (0.230)   & 0.270 (0.739)       & 0.505 (1.496)     & 0.222 (0.450)      \\
 $\gamma_{A},\gamma_{B}>0,\:\:\gamma_{C},\gamma_{D}<0  $     &  50\%   & 199 $\pm$12      &   flat       &    0.042 (0.090)   & 0.121 (0.301)       & 0.240 (0.643)     & 0.069 (0.139)    \vspace{3mm}
 \\

Shear:                                                       &  0\%    & 270 $\pm$10      &   flat       &  0.107 (0.244)     & 0.293 (0.792)       &  0.553 (1.638)    & 0.222 (0.450)      \\
$\gamma_{A},\gamma_{B}<0,\:\:\gamma_{C},\gamma_{D}>0  $      &  50\%   & 181 $\pm$4       &   flat       &  0.044 (0.093)     & 0.135  (0.330)       & 0.278 (0.729)     & 0.069 (0.239)     \\ \hline

\end{tabular}

\end{center}

\vspace{5mm}
 $v_{stream}=233\,km\,sec^{-1}$ 
\begin{tabular}{|c|c|c|c|c|c|c|c|}
\hline
 Trajectory                                                  & Smooth  & Equiv.           & Bayesian       & $m_{low}(99\%)$  & $m_{low}(95\%)$  & $m_{low}(90\%)$ &  mode     \\ 
Orientation                                                  & Matter  & Vel. $(km\,sec^{-1})$         & prior         &   ($M_{\odot}$)  &  ($M_{\odot}$)    &  ($M_{\odot}$) &  ($M_{\odot}$)    \\ \hline\hline 
Shear:                                                       &  0\%       & 129 $\pm$7       &   log      & 0.010 (0.023)      & 0.024 (0.061)      & 0.039 (0.113)     & 0.027 (0.043)      \\ 
 $\gamma_{A},\gamma_{B}>0,\:\:\gamma_{C},\gamma_{D}<0  $     &  50\%      &  99 $\pm$5       &   log      & 0.006 (0.011)      & 0.013 (0.030)      & 0.022 (0.056)     & 0.013 (0.021)     \vspace{3mm} 

  \\
Shear:                                                       &  0\%       & 120 $\pm$2       &   log      & (0.011 (0.023)      & 0.025 (0.064)       & 0.042 (0.120)   & 0.021 (0.043)      \\ 
$\gamma_{A},\gamma_{B}<0,\:\:\gamma_{C},\gamma_{D}>0  $      &  50\%      &  85 $\pm$3      &    log      & 0.005 (0.009)      & 0.011 (0.024)       & 0.018 (0.044)   & 0.010 (0.021)      \\ \hline 
Shear:                                                       &  0\%       & 129 $\pm$7       &   flat     & 0.028 (0.071)      & 0.105 (0.319)      & 0.251 (0.823)     & 0.034 (0.069)      \\ 
 $\gamma_{A},\gamma_{B}>0,\:\:\gamma_{C},\gamma_{D}<0  $     &  50\%      &  99 $\pm$5       &   flat     & 0.014 (0.031)      & 0.053 (0.144)     & 0.136 (0.395)     & 0.017 (0.027)     \vspace{3mm} 

  \\
Shear:                                                       &  0\%       & 120 $\pm$2       &   flat     & 0.027 (0.069)      & 0.108 (0.33)        & 0.273 (0.897)   &  0.034 (0.687)     \\ 
$\gamma_{A},\gamma_{B}<0,\:\:\gamma_{C},\gamma_{D}>0  $      &  50\%      &  85 $\pm$3      &    flat     & 0.014 (0.031)      & 0.060 (0.159)       & 0.162 (0.454)   &  0.017 (0.027)     \\ \hline 

\end{tabular}
\vspace{5mm}

\end{center}
\end{table*}

In this section we describe a method for determining the lower limit for the average compact object mass. The microlensing rate has a minimum possible value determined by the size of the stellar velocity dispersion, thus there is a minimum mass which can explain the observed rate. This is quantified by noting that the measured effective transverse velocity must be greater than the equivalent transverse velocity.

 There is a simple scaling that relates models of the same optical depth but consisting of model stars with a different mass. The dimensionless Einstein radius of a point mass is $\sqrt{m}$. Therefore, reducing all microlens masses by a factor $a$ reduces all physical distances (in both the source and lens planes) between dimensionless coordinates by a factor $\sqrt{a}$ (eg. Witt, Kayser \& Refsdal 1993). For a given transverse velocity this results in an increase by a factor of $1/\sqrt{a}$ in the gradient of the light-curve at all points. Similarly, the rate of change of magnification at a static source point due to stellar proper motions increases by a factor $1/\sqrt{a}$. Since it results from the proper motions the minimum rate is larger if a smaller mean microlens mass is assumed. However the minimum rate cannot statistically exceed the observed rate. The ratio of equivalent transverse velocity to velocity dispersion is independent of the mean microlens mass, however the measurement of effective transverse velocity is proportional to $\sqrt{a}$. This allows limits to be placed on the value of $a$ and therefore the stellar mass from the observed microlensing rate.

\begin{figure}
\vspace*{155mm}
\includegraphics{fig3.epsi}
\caption{\label{mass_lim} The differential distributions for the mean microlens mass. The solid and dotted curves represent the resulting functions when the photometric error was assumed to be $\sigma_{SE}$=0.01 mag in images A/B and $\sigma_{SE}$=0.02 in images C/D, and $\sigma_{LE}$=0.02 mag in images A/B and $\sigma_{LE}$=0.04 in images C/D. The light and dark lines correspond to the assumptions of logarithmic and flat priors for effective transverse velocity referred to in the text.}
\end{figure}

\begin{figure*}
\vspace*{130mm}
\includegraphics{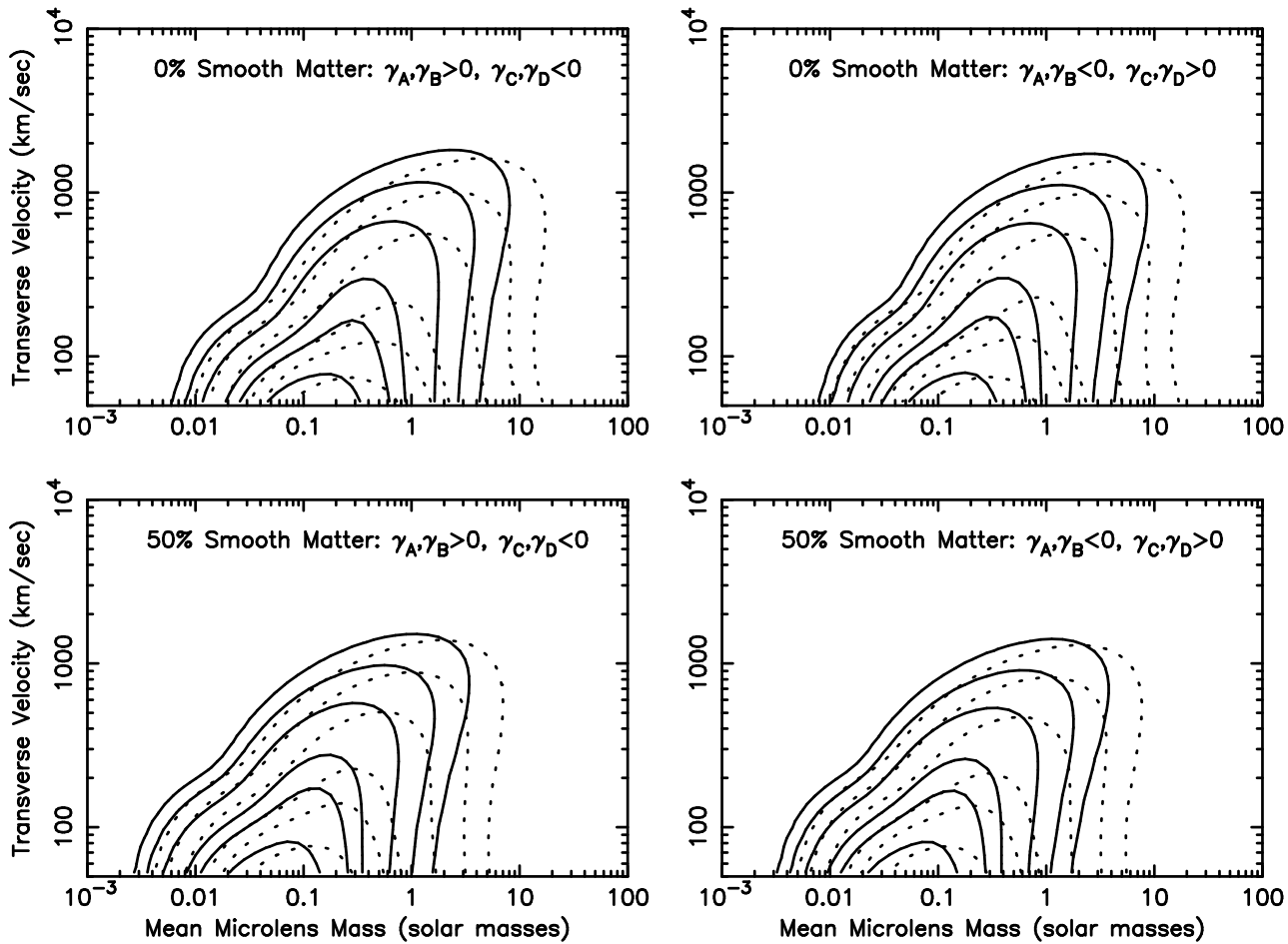}
\caption{\label{contour}Contours of percentage peak height of the function $p_{m,v}(\langle m\rangle,V_{tran})$. The contours shown are the 0.1\%, 1.0\%, 3.6\%, 14\%, 26\% and 61\% levels. The solid and dotted curves represent the resulting functions when the photometric error was assumed to be $\sigma_{SE}$=0.01 mag in images A/B and $\sigma_{SE}$=0.02 in images C/D, and $\sigma_{LE}$=0.02 mag in images A/B and $\sigma_{LE}$=0.04 in images C/D.}
\end{figure*}

\subsection{\label{propmot}Mass limits from random proper motions}

 For a model where $\gamma_{A},\gamma_{B}>0$, and the simulated errors are assumed to be $\sigma_{SE}=0.01$ and $\sigma_{SE}=0.02$ mags in images A/B and C/D respectively, the effective transverse velocity is less than $1346\sqrt{\langle m\rangle} \,km\,sec^{-1}$ at the 99\% level, where $\langle m\rangle$ is the microlens mass assumed.
 In the absence of a galactic transverse velocity, the minimum level of microlensing is described by the equivalent transverse velocity of the stellar velocity dispersion $V_{equiv}(DISP)=280\,km\,sec^{-1}$ (see Tab.~\ref{stream_tab}). The value of $\langle m\rangle$ at which $V_{upper}(99\%)$ drops below $V_{equiv}(DISP)$ is $m_{low}(99\%)$:
\begin{equation}
\label{mass_low}
m_{low}(99\%)=\left(\frac{V_{equiv}(DISP)}{V_{upper}(99\%)}\right)^{2}\sim 0.04. 
\end{equation}
\noindent The minimum mass of stars in this model is therefore $\sim 0.04M_{\odot}$. This calculation assumes a value for the one dimensional velocity dispersion of $165\,km\,sec^{-1}$. However we note that all lower limits obtained in this section are proportional to the square of this value. The procedure also assumes that the measured effective transverse velocity is independent of the assumed source size. WWTb show that this is approximately true for source sizes smaller than $\approx 4\times 10^{16}\sqrt{m}\,cm$. This value compares favourably with the findings of Wambsganss, Paczynski \& Schneider (1990) that a source smaller than $\sim 2\times 10^{15}\sqrt{\langle m\rangle /.225}\,cm$ is required to typically reproduce the observed HME amplitude.

The curves shown in Fig.~\ref{vel_lim} represent the cumulative probability $P_v(V_{gal}<V_{eff}\sqrt{m})$ that the effective transverse velocity $V_{gal}$ is less than $V_{eff}\sqrt{m}$.
It follows that the cumulative probability that the mean microlens mass is less than the value 
\begin{equation}
\nonumber M=\left(\frac{V_{equiv}(DISP)}{V_{eff}}\right)^{2}
\end{equation}
is
\begin{equation}
\nonumber U_m(\langle m\rangle<M|V_{equiv})=1-P_v(V_{gal}<V_{eff}\sqrt{M}).
\end{equation}

The calculation of $m_{low}(99\%)$ performed above assumes that the physical transverse velocity of the galaxy with respect to the observer source line of sight $V_{tran}=0$. However since the galaxy may have a component of transverse motion, $V_{equiv}$, which describes the minimum microlensing rate is a function of $V_{tran}$. The dependence of $V_{equiv}$ on $V_{tran}$ is determined according to the relationships described in WWTb. Since the galactic transverse velocity is not necessarily zero, $U_m(\langle m\rangle<M|V_{equiv})$ must be determined at a series of transverse velocities and combined with prior probabilities for $V_{tran}$ to obtain estimates for the probability of the mean microlens mass. We have assumed the flat and logarithmic priors employed for the calculation of $P_v(V_{gal}<V_{eff})$:
\begin{eqnarray}
\nonumber U_m(\langle m\rangle<M)&=&\\
&&\hspace{-25mm}\int U_m(\langle m\rangle<M|V_{tran})p_{prior}(V_{tran})dV_{tran}
\end{eqnarray}
The probability density $p_{m}(\langle m\rangle)$ for the mean microlens mass is obtained by taking the derivative $\frac{d\,U_m}{d\,\langle m\rangle}$. Note that this distribution does not represent the mass function.

Fig.~\ref{mass_lim} shows the functions $p_{m}(\langle m\rangle)$ for the 0\% smooth matter and 50\% smooth matter models for both directions considered.
The solid and dotted curves represent functions assuming the photometric error to be $\sigma_{SE}$=0.01 mag in images A/B and $\sigma_{SE}$=0.02 in images C/D, and $\sigma_{LE}$=0.02 mag in images A/B and $\sigma_{LE}$=0.04 in images C/D.
 The light and dark lines refer to the logarithmic and flat priors for transverse velocity. Results for $m_{low}(99\%)$, $m_{low}(95\%)$, $m_{low}(90\%)$ and the mode of the distribution are shown in the top panel of Tab.~\ref{stream_tab}. In each column, the values correspond to the two assumptions for the simulated error referred to above. $m_{low}(99\%)$ is $>0.02M_{\odot}$ in all models considered. The mode of the distribution ranges between 0.04$M_{\odot}$ and 0.45$M_{\odot}$. It is interesting to note that these models produce distribution modes that are similar to the mean of a Salpeter distribution with a lower cutoff at the hydrogen burning limit. We also note that the distribution is highly asymmetric, with the mode being of the same order as the lower limit.

\subsection{The effect of systematic assumptions}

 In contrast to the calculation of $P_v(V_{gal}<V_{eff}\sqrt{\langle m\rangle})$ the assumed prior for $V_{eff}$ makes a significant difference to the microlens mass limits obtained. In particular, low mass microlenses are less likely if large transverse velocities are assumed. This dependence on the prior illustrates the limitation of the light curve data for breaking the degeneracy between galactic transverse velocity and microlens mass in the high mass-velocity regime.

 The assumption of a larger photometric uncertainty yields a lower estimate of the transverse velocity (WWTb). This in turn means that a larger estimate is made for the mean microlens mass. The effect is readily apparent in Fig.~\ref{mass_lim} as well as from Tab.~\ref{stream_tab}.

We find lower microlens mass limits if a non-zero fraction of smooth matter is assumed. This results from the combination of a larger determination of effective transverse velocity and a smaller relative contribution to microlensing of the stellar proper motions resulting in a smaller minimum for microlensing rate. This is expected since no microlensing will be observed due to a galaxy composed entirely of continuously distributed matter. The plots in Fig.~\ref{mass_lim} demonstrate the level of dependence on smooth matter component.

$p_{m}(\langle m\rangle|V_{tran})$ is extremely sensitive to the level of photometric error assumed at large $\langle m\rangle$. In addition, microlenses of arbitrarily high mass can produce the required rate in combination with a correspondingly large transverse velocity. Therefore at large $\langle m\rangle$, $p_{m}(\langle m\rangle)$ is a sensitive function of the prior assumed. For these reasons our method does not place useful upper limits on the mean microlens mass.

\subsection{Simultaneous limits on mean microlens mass and transverse velocity}
\label{simul}

To explore the co-dependence of the determined mean microlens mass and galactic transverse velocity we have computed the two dimensional distribution 
\begin{equation}
p_{m,v}(\langle m\rangle,V_{tran})=p_{m}(\langle m\rangle|V_{tran})\,p_{v}(V_{tran}|\langle m\rangle).
\end{equation}
Figure~\ref{contour} shows contour plots of this distribution for the assumed source orientations, smooth matter fraction and photometric errors. All possible combinations of the two values of each parameter are included, so that eight models are shown. The contours are at 0.1, 1.0, 3.6, 14, 26, 61 percent of the peak height, and so the extrema of the inner 4 contours represent $4\sigma$, $3 \sigma$, $2\sigma$, and $1\sigma$ limits on the single variables. The most important benefit of combining the distributions of $V_{tran}$ and $\langle m \rangle$ is that upper limits can be placed on the mean microlens mass and galactic transverse velocity. The limits on $\langle m \rangle$ and $V_{tran}$ are subject to both random and systematic uncertainties. Assuming that the chosen values for these parameters bracket the likely models, upper and lower limits for $\langle m \rangle$, and an upper limit for $V_{tran}$ can be found by using the most extreme values of the contour limits. This yields the 95 per cent result that $0.01 M_{\odot} \la \langle m \rangle \la 1 M_{\odot}$, and $V_{tran}\la 300\,km\,sec^{-1}$ (95\%). The upper limit is lower than that obtained by Lewis \& Irwin (1996), although sufficiently high that it does not conflict with any popular models of galactic halos or bulges. The elongation of the outer contours with a power law slope of $\frac{1}{2}$ illustrates the degeneracy between microlens mass and transverse velocity at large mass.

\subsection{The effect of trajectory direction}

\begin{figure}
\vspace*{85mm}
\includegraphics{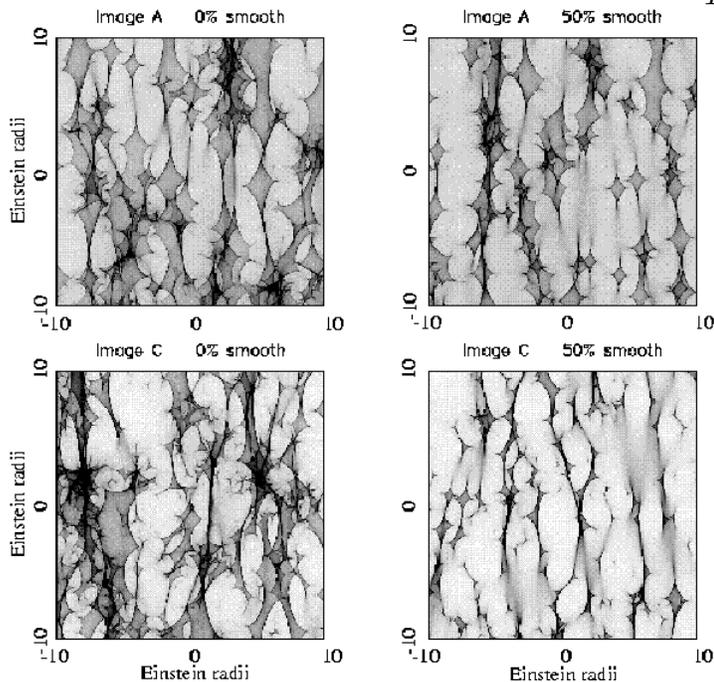}
\caption{\label{magmaps} The magnification maps for images A (top) and C (bottom), in the cases of 0\% smooth matter (left) and 50\% smooth matter (right). The point masses each had a mass of 1$M_{\odot}$}
\end{figure}

 Fig.~\ref{magmaps} shows the magnification maps for images A and C in the cases of 0\% and 50\% continuously distributed matter. The caustic networks in both images are significantly stretched so that a source could travel a significant distance in the direction of caustic clustering and be subject to little microlensed flux variation. 
The measured effective transverse velocity is dependent on trajectory direction, with a larger measurement resulting from cases where the source trajectory is perpendicular to the shear in images C and D ($\gamma_{C},\gamma_{D}<0$). The larger measurement quantifies the greater stretching of the caustic network in these images, creating a larger difference between the rate of microlensing for sources moving with and against the shear. The variation with orientation of the measurements of effective transverse velocity is slightly larger when the model contains a component of smooth matter due to additional stretching of the caustic network.

A similar argument applies to the determination of the equivalent transverse velocity of a stellar velocity dispersion. The rate of microlensing due to stellar proper motions at a series of fixed source points is independent of the direction in which those points are sampled, regardless of the contributing fraction of smoothly distributed matter. However microlensing that results from a transverse velocity has a rate that varies more with direction in images C and D. The equivalent transverse velocity is therefore a function of direction. The effect is more pronounced in models that include a smoothly distributed mass component.

Thus there are two separate effects. Firstly, if the source trajectory is aligned with the image C-D axis the measured effective transverse velocity increases. This effect is magnified by the inclusion of a component of smooth matter. In addition, the equivalent transverse velocity of the stellar proper motions is also increased under the same circumstances. Interestingly, during our calculation of lower mass limit these effects tend to have a cancelling effect, rendering the limits reasonably insensitive to the source trajectory direction assumed.

\subsection{Mass limits from stellar stream motions}

The calculations in Sec \ref{propmot} and Sec \ref{simul} assumed that microlensing results from the combination of stellar proper motions and galactic transverse velocity. However the bulge dynamics may be dominated by rotational motion resulting in stream motions of the starfield. For an isothermal sphere model of the bulge, the rotational velocity is $\sqrt{2} \sigma_{*}=233\,km\,sec^{-1}$.
Kundic, Witt \& Chang (1993) obtained the expression for the caustic velocity $v_{caust}$ resulting from a stream motion $\vec{v}_{stream} = (v_{stream},0)$:
\begin{equation}
\label{stream}
v_{caust}=(1-\gamma-\kappa_{c}) |\vec{v}_{stream}|.
\end{equation}
 If the rotational motion is assumed to be circular then the stream motions are perpendicular to the image-galactic centre axis, and therefore parallel to the shear vector at each image. We choose the stream motions to be in the $x_{1}$ direction making the shear positive for all 4 images. An equivalent transverse velocity was defined for the stream motion and determined in analogy to that for the stellar proper motions. For a circular stream motion of $v_{stream}=233\,km\,sec^{-1}$ equivalent transverse velocities ($V_{equiv}(STREAM)$) were calculated for the two transverse directions discussed previously. These values are presented in Tab.~\ref{stream_tab}. When combined with the equivalent velocities for an isotropic velocity dispersion, the values describe the contribution to the microlensing rate of the extreme cases for stellar motions in the galactic bulge.

The lower panel in Tab.~\ref{stream_tab} shows results for models with stream motions. Values are shown for $m_{low}(99\%)$, $m_{low}(95\%)$, $m_{low}(90\%)$ and the mode for all models considered. $m_{low}(99\%)$ is $> 0.005M_{\odot}$ in all models.
The equivalent transverse velocities presented in Tab.~\ref{stream_tab} show that an orbital stream motion produces a much lower microlensing rate than a transverse velocity of equal magnitude (independent of its direction). The reason for this behaviour is apparent from Eqn. \ref{stream}; a positive shear reduces the caustic velocity resulting from stream motions. In addition, the caustic velocity and hence microlensing rate are also decreased when there is a component of smooth matter. The lower microlensing rate translates to an equivalent transverse velocity that is smaller than the one obtained from a stellar velocity dispersion. This in turn means a lower estimate of $m_{low}(P)$. In the case discussed, the mass limit obtained is only $\sim 0.2$ times that where the proper motions are assumed to be isotropic.

The last row of Tab.~\ref{ml} summarises the range of values for the mean microlens mass obtained by this study. These compare favourably with the results of related studies, none of which find that Jupiter mass compact objects are the dominant component of mass in the bulges and or halos of the spiral galaxies studied. 

\section{Conclusion}

Calculations of the transverse velocity from published monitoring data of Q2237+0305, as well as complimentary characterisations of caustic clustering from previous studies show that the caustic networks produced by populations of microlenses have a structure independent of the mass spectrum, but a scale length proportional to the square root of the mean microlens mass. Previous determinations of average mass from microlensing data have therefore been proportional to the square of the unknown transverse velocity.
 However a minimum rate of microlensing is produced by the proper motions of microlenses in the bulge of the lensing galaxy. We have used limits on the effective transverse velocity obtained from published monitoring data for Q2237+0305 in combination with calculations of the microlensing effect of isotropic stellar proper motions to calculate the probability for the mean microlens mass. We find that the most likely value for the mean microlens mass is 0.04-0.45$M_{\odot}$. The lower limits on the mean microlens mass responsible for the observed microlensing are $\langle m\rangle \,> 0.02-0.24M_{\odot}$ (99\% level). A similar calculation assuming an orbital stream motion of stars rather than a velocity dispersion produces a most likely value of 0.01-0.05$M_{\odot}$ and a lower limit of $\langle m\rangle \,>0.005-0.071M_{\odot}$ (99\% level). Through joint consideration of the probability for mean microlens mass and transverse velocity under the assumption of an isotropic velocity dispersion we obtain a mass between $0.01M_{\odot}$ and $1.0M_{\odot}$ (95\%), and a galactic transverse velocity less than 300$\,km\,sec^{-1}$ (95\%).

Our result that very low mass objects do not comprise a significant mass fraction of microlenses in Q2237+0305 supports the results of Schmidt \& Wambsganss (1998) who rule out microlenses in Q0957+561 having $\langle m\rangle~\ll~0.01\left(\frac{v_t}{600\,km\,sec^{-1}}\right)^2$. The most likely mean microlens mass of $\sim 0.2M_{\odot}$ supports both the conclusions of Lewis \& Irwin (1996) who found that microlenses in Q2237+0305 have $0.1M_{\odot}\la\langle m\rangle\left(\frac{v_t}{600\,km\,sec^{-1}}\right)^2\la10M_{\odot}$, and of the MACHO collaboration (Alcock et al. 1997a,b) who find from microlensing in the Milky Way that $\langle m\rangle\ga 0.05M_{\odot}$ in the halo and $0.1M_{\odot}\la\langle m\rangle\la1.0M_{\odot}$ towards the bulge.

 While our limits depend on the correctness of the published macrolensing parameters, as well as on the fraction of smoothly distributed matter assumed, they do not depend on other unknowns present in the problem including the source size, source intensity profile, effective transverse velocity and direction of the source trajectory. We have investigated models having smooth matter fractions of 0 and 0.5, and find that smaller masses can explain the observed microlensing rate if a non-zero smooth matter fraction is assumed. Our results depend on the Bayesian prior chosen for galactic transverse velocity. However we have chosen priors to bracket physically reasonable choices and find that the contribution of the prior to the systematic uncertainty is smaller than that of the assumption for smooth matter fraction.

Our results unambiguously rule out a significant contribution of Jupiter mass compact objects to the mass distribution of the bulge whether the microlenses move predominantly in orbital or random motions. However if the microlens proper motions are distributed according to an isotropic velocity dispersion then the lower limit obtained is of the same order as the hydrogen burning limit.

\section*{Acknowledgements}

The authors would like to thank Chris Fluke and Daniel Mortlock for helpful discussions. This work was supported in part by NSF grant AST98-02802. JSBW acknowledges the support of an Australian Postgraduate award and a Melbourne University Overseas Research Experience Award.

\label{lastpage}

\end{document}